\documentclass[aip,jcp,preprint]{revtex4-1}

\usepackage{bm} 

\usepackage{graphicx}

\begin{document}


\title{
Fast convergence to equilibrium for long-chain polymer melts 
using a MD/continuum hybrid method
}


\author{Yasuhiro Senda}
\author{Miyuki Fujio}
\author{Shuji Shimamura}
\affiliation{Department of Applied Science, Yamaguchi University, Yamaguchi, 755-8611, Japan}
\author{Janne Blomqvist}
\author{Risto M Nieminen}
\affiliation{Department of Applied Physics, Aalto University, School
  of Science, P.O. Box 11100, FI-00076 Aalto, Finland}




\begin{abstract}
Effective and fast convergence toward an equilibrium state for
long-chain polymer melts is realized by a hybrid method coupling
molecular dynamics and the elastic continuum.  The required simulation
time to achieve the equilibrium state is reduced drastically compared
with conventional equilibration methods.  The polymers move on a wide
range of the energy landscape due to large-scale fluctuation generated
by the elastic continuum.  A variety of chain structures is generated
in the polymer melt which results in the fast convergence to the
equilibrium state.

\end{abstract}


\maketitle


\section{Introduction}
Atomistic simulations for polymers have been studied intensively.
Especially molecular dynamics (MD) calculations for polymers in order
to reveal the dynamical behavior of the chain structure have been
carried out.  MD calculations for long-chain polymers, however, have
been limited by the massive computational costs due to the very long
relaxation times of entangled long-chain polymer melts.  According to
reptation theory, the relaxation time of an entangled polymer melt
consisting of chains with $N$ monomers scales as $N^{3}$ .  This means
that a prohibitively long simulation time is needed to relax a dense
polymer melt.  Moreover, a complex system such as an entangled polymer
melt exhibits a huge number of local-minimum energy states in the free
energy surface.  Energetic barriers much larger than the thermal
energies separate the initial configurations from the final
equilibrium states, which leads to relaxation times far greater than
currently accessible computational resources allow.

The coarse-grained (CG) approach for polymer molecules, in which
multiple atoms are combined into a large bead, enable us to extend the
spatial and time scales of the simulation.  In particular, the time
scales up to several orders of magnitude from the atomistic level
\cite{kremer1990}.  The length of the chain polymer is, however,
limited to the order of $10^{4}$ monomers even with the CG approach
\cite{putz2000}.
  
A variaty of method has been proposed for obtaining well-equilibrated CG polymer.  
Auhl {\it et al.} used the initial configuration reducing the density fluctuation
and a double-pivot algorithm\cite{karayiannis2002} for a MD calculation\cite{auhl2003}. 
They demonstrated the effectiveness of their method in long-chain polymer simulations.  
Gao proposed a method of polymer chain genration by 
connecting the polymer to monomers, combining with the relaxation of polymer 
comformations by MD step\cite{gao1995}. 
Perez  {\it et al.} confirmed that the relaxation is performed while the chains are generated and 
showed an applicability of this method to complex polymers such as nanostructued polymer\cite{perez2008}.
Subramanian generated the well equilibrited polymer by affinely scaling the simulation box and 
adding the beads along the contour of the chain, and applied it to the cyclic polymers \cite{subramanian2010}.
Methods for overcoming the local energy minimum on the energy surface have been intensively
studied \cite{voter1997, sorensen2000,
  jonsson1998,passerone2001,rosso2002}.  The multicanonical MD method
\cite{nakajima1997} enables sampling over a much larger phase space,
and was applied to a CG model of protein folding \cite{isobe}.
 
In the present paper, MD simulations for long-chain polymer melts are
performed by a hybrid MD/continuum method, in which the dynamics of
the atoms is coupled with the those of the continuum degrees of
freedom concurrently. This hybrid method was originally proposed by
our group, and it has been applied to a simple one-dimensional system
\cite{kim2007,senda2009}, in which the spring force of the continuum
acts on the atomic chain system and generates large-scale fluctuations
and a variety of atomic phonon modes in the atomic chain.  In the
present paper, the hybrid method is applied to a polymer melt
consisting of long-chain polymers.  We demonstrate that the
large-scale fluctuation induces a large number of states of the
long-chain polymers, which overcomes the energetic barriers, and leads
to the fast convergence towards the final equilibrium state.  Our
purpose is to show the result of accelerating MD calculations using
the MD/continuum hybrid method and the effectiveness of this method
for long-chain polymer simulations.

\section{Method}

We describe a single polymer as a bead-spring chain in which monomers
of the polymer are represented by spherical beads.  The beads have an
excluded volume described by the repulsive force of the 12-6 Lennard
Jones potential.
\begin{equation}
U_{LJ}(r)=
\left\{
 \begin{array}{lll}
       4 \epsilon \{(\frac{\sigma}{r})^{12}-(\frac{\sigma}{r})^{6}+\frac{1}{4}\}& & r \le r_{c} \\
       0                               & & r >  r_{c}
 \end{array}
\right.,
\end{equation}
where the cutoff radius $r_{c}$ is set as $2^{1/6} \sigma$.  Each
bead is connected with the neighboring beads in the polymer chain via
a finite extensible non-linear elastic (FENE) potential as
\begin{equation}
U_{FENE}(r)=
\left\{
 \begin{array}{lll}
      -0.5kR_{0}^{2}ln(1-(r/R_{0}^{2})^{2}) & & r \le R_{0} \\
       \infty                               & & r >  R_{0}
 \end{array}
\right.,
\end{equation}
where $k=30\epsilon/\sigma^{2}$ and $R_{0}=1.5 \sigma$.  In
addition, we adopt a bending potential for the polymer defined by
\begin{equation}
U_{bend}(\theta)=k_{\theta}(1-cos\theta)\ ,
\end{equation}
where $\theta$ is the angle between the neighboring bonds within the
polymer chain. $k_{\theta}$ is set to 0.25 $\epsilon$. 
This bending potential is applied in order to illustrate the wide range energy covered by the hybrid method
as will be mentioned in disccusion.   
The calculation is performed using the software package ESPResSo \cite{limbach2006}.

\begin{figure}
\begin{center}
\includegraphics[width=0.7\columnwidth]{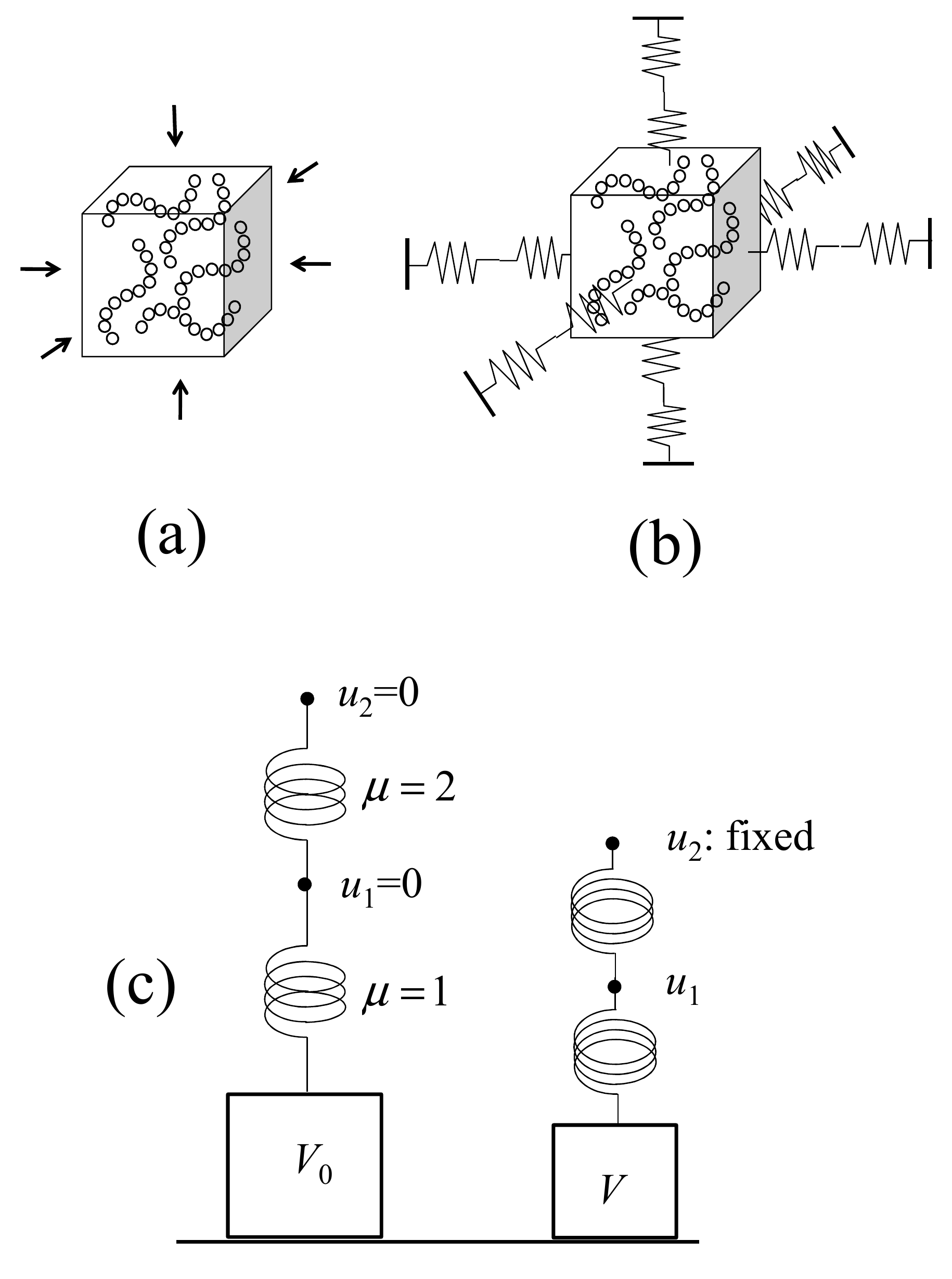}
\end{center}
\caption{ Schematic views of (a) a standard constant-pressure MD model
  and (b)(c) the hybrid model.  (a) The polymer system consists of monomers
  (open circle) in a cubic MD cell under an external
  constant pressure.  (b) The springs enclose the cubic MD cell and
  isotropic forces by the springs act on the polymer system. 
  (c) Schematic image of the hybrid model. On left-hand side no displacemnent is applied on springs, 
while on the right-hand, MD cell is compressed by the springs}
\label{hybrid}
\end{figure}

The above CG polymer model is connected with the elastic continuum.
Details of the MD/continuum hybrid method are explained in
Ref. \cite{kim2007}, and here the procedure of the hybrid method
is only briefly explained.  The elastic continuum surrounds the MD cell
including the CG polymer model and the elastic stress acts on the MD
cell.  In the case of a constant-pressure MD method
\cite{andersen1980}, the constant pressure acts on the MD cell.  The
constant pressure is replaced by the elastic stress of the continuum
in the hybrid method as shown in Fig.\ref{hybrid}.  Since the present
system is considered to be isotropic, the cubic MD cell is under
isotropic stress of the elastic continuum, which is described by the
springs as shown in Fig. \ref{hybrid}.  According to the procedure of
the MD/continuum hybrid method \cite{kim2007}, we can describe the
Lagrangian functional $L$ of the present hybrid model consisting of $N$
particles in the MD cell with the volume $V$ and the $N_{s}$ springs as

\begin{widetext}
\begin{eqnarray}
\label{eq??}
  L(\{\bm{s}_i ,\dot {\bm{s}}_i \},V,\dot {V},\{u_\mu ,\dot {u}_\mu  \})=
\sum\limits_{i=1}^N {\frac{m V^{2/3} {\dot {\bm{s}}_i}\cdot{\dot {\bm{s}}_i}} {2}} 
-\phi (\{ \bm{s}_{i}\},V) 
\nonumber \\
+\frac{Q\dot {V}^2}{2} 
-\frac{K}{2}(V-V_{0}-u_{1})^2
+\frac{M\dot{u_{1}}^{2}}{2}
\nonumber \\
+\sum\limits_{\mu=2}^{N_{s}} 
\frac{M\dot {u}_{\mu}^2}{2} 
-\frac{K (u_{\mu -1} -u_\mu )^2}{2}
, 
\end{eqnarray}
\end{widetext}
where $\bm{s}_{i}$ are the scaled coordinates of the particles, such that the Cartesian positions $\bm{r}_{i}$ are $\bm{r}_{i}=V^{1/3}\bm{s}_{i}$. $u_{\mu}$ are the reduced displacements of the springs in volume units.  $m$ and
$M$ are the masses of the particles and the springs, and $Q$ is the inertial mass
for the motion of the volume $V$, which is also presented in the standard
constant-pressure MD method \cite{andersen1980}.  $\phi$ is the
potential energy between the particles of the CG polymer model. 
The fourth term $\frac{K}{2}(V-V_{0}-u_{1})^2$ corresponds to the elastic potential
energy of the first spring ($\mu=1$) as is illustrated in Fig.1(c). 
This energy depends on the volume $V$, the displacement of this spring $u_{1}$, 
in which $K$ is the spring constant. 
$V_{0}$ corresponds to the volume under no displacement appiled on springs.
The initial displacements and their velocities of the springs $u_{\mu},\dot {u}_{\mu}$ 
are set so as to apply the pressure on the CG model. 
The displacement of terminal spring $u_{Ns}$ is fixed.   
The equations of the motion for the particles, volume and the springs are derived
easily from the above Lagrangian.

\begin{equation}
\label{eq7}
 m \frac{d \dot{\bm{s}}_{i} }{dt}=-V^{-2/3}\frac{\partial \phi }{\partial \bm{s}_i } 
- \frac{2}{3} \frac{\dot{V}}{V} m \dot{\bm{s}}_{i},
\end{equation}

\begin{widetext}
\begin{equation}
\label{eq8}
Q\frac{d\dot {V}}{dt}=\frac{1}{3V} 
\sum\limits_{i=1}^N 
{ \left( m V^{2/3}\dot {\bm{s}}_i^2 - \bm{s}_{i}\cdot \frac{\partial \phi}{\partial \bm{s}_{i}} \right) }
-K(V-V_{0}-u_{1}),
\end{equation}

\begin{equation}
\label{eq9}
M \frac{d\dot {u}_\mu }{dt}=\left\{ {{\begin{array}{*{20}c}
 {K (V-V_0 -u_\mu )-K (u_\mu -u_{\mu +1} )} \hfill \\
 {K (u_{\mu -1} -u_\mu )-K (u_\mu -u_{\mu +1} )} \hfill \\
\end{array} }{\begin{array}{*{20}c}
 , \hfill \\
 , \hfill \\
\end{array} }} \right.{\begin{array}{*{20}c}
 \hfill & {(\mu =1)} \hfill \\
 \hfill & {(\mu =2,3,\cdots,N_{s} -1 )}. \hfill \\
\end{array} }
\end{equation}
\end{widetext}
In the equation of motion (\ref{eq8}) for the volume $V$, the first
two terms on the right correspond to the internal pressure of the CG
polymer system, and the last term is the elastic force caused by the
adjacent spring ($\mu=1$).  This equation plays a role for connecting
the CG polymer system of the MD cell to the springs.  The simultaneous
equation of the degrees of $\{\bm{s}_{i}\}$, $V$ and $\{u_{\mu}\}$ is
integrated numerically and we obtain the time convolution of the
coordinations of the polymer, the volume of the MD cell, and the
displacement of the springs.

In our simulations, we use a CG polymer consisting of 400 beads and we
place 10 such CG polymers in the MD cell.  The total number of beads
is thus 4000.  The number density of the CG polymer liquid is set to
0.85 $\sigma^{-3}$ and the temperature is set to 1.0 $\epsilon/k_{B}$.
The time unit of the calculation is $\tau=\sigma(m/\epsilon)^{1/2}$.
The integration of the equation of the motion is performed using a
time step 0.006 $\tau$.

To show the effectiveness of the hybrid method, we investigate the
required simulation time to achieve the equilibrium state.  As an
initial configuration for the MD calculation, each CG polymer is set
to have a long stretched chain structure.  Conventional Andersen
constant-pressure MD is performed using the same initial configuration
in order to compare it to the result of the hybrid method.


\section{Results}

The single-chain structure is characterized by the end-to-end distance
$R$ of a single-chain polymer.  
%
Using $k_{\theta}=0.25 \epsilon$ of the bending potential and the average bond length $<b>=0.97 \sigma $, 
the square root of $<R^{2}>$ is derived to be 26.9 $\sigma$ \cite{auhl2003}.

\begin{figure}
\begin{center}
\includegraphics[width=0.6\columnwidth]{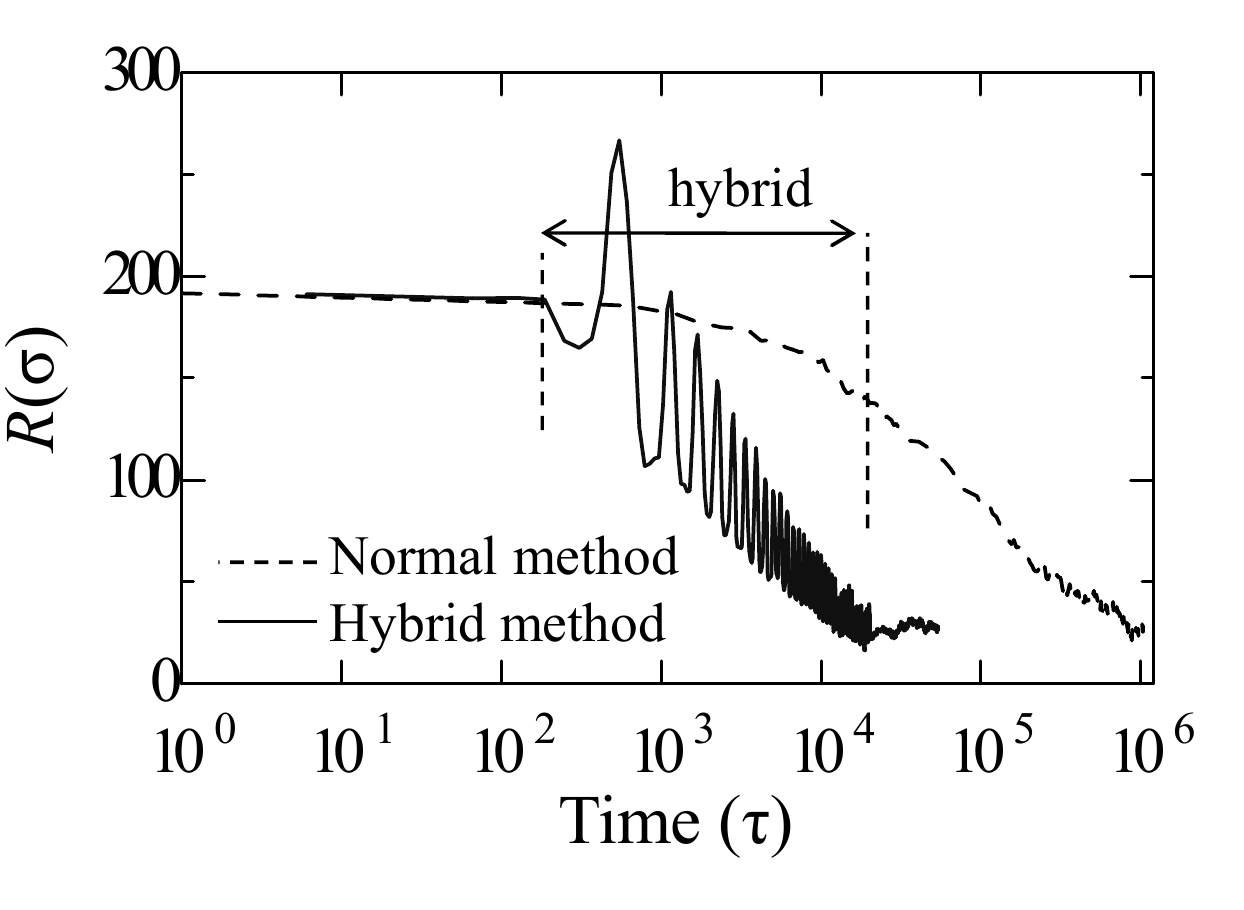}
\end{center}
\caption{ Time convolutions of the end-to-end distance $R$ of polymer melts 
obtained by the conventional method (broken line) and hybrid method (solid line).  
The calculation with the hybrid method stops at $t=2.0 \times 10^{4} \tau $, and the
  constant-pressure MD calculation is continued after that.  }
\label{endtoend}
\end{figure}

The time convolutions of the calculated $R$ in the present MD
calculations are monitored in Fig. \ref{endtoend}, where the result
obtained by the conventional method is compared with that from the
hybrid method.  In the conventional MD method, considerably longer
simulation time ($t \sim 1.0 \times 10^6 \tau $) is needed to obtain a
stable value of $R$, which is in good agreement with the above
analytical value.  The long-chain polymer such as the present model
has very slow diffusion time and requires a long simulation time to
reach the relaxation using a conventional MD method.  In contrast, the
$R$ of the hybrid method is fluctuating wildly and rapidly converges
toward the equilibrium value.  After a simulation with the hybrid
method until $t=2.0\times 10^{4} \tau $, the calculation is continued
with the conventional MD method.  The required time to reach the
equilibrium state using the hybrid method is $t \sim 10^{4} \tau $,
which is about one hundredth of the time required using the
conventional method.

\begin{figure}
\begin{center}
\includegraphics[width=0.8\columnwidth]{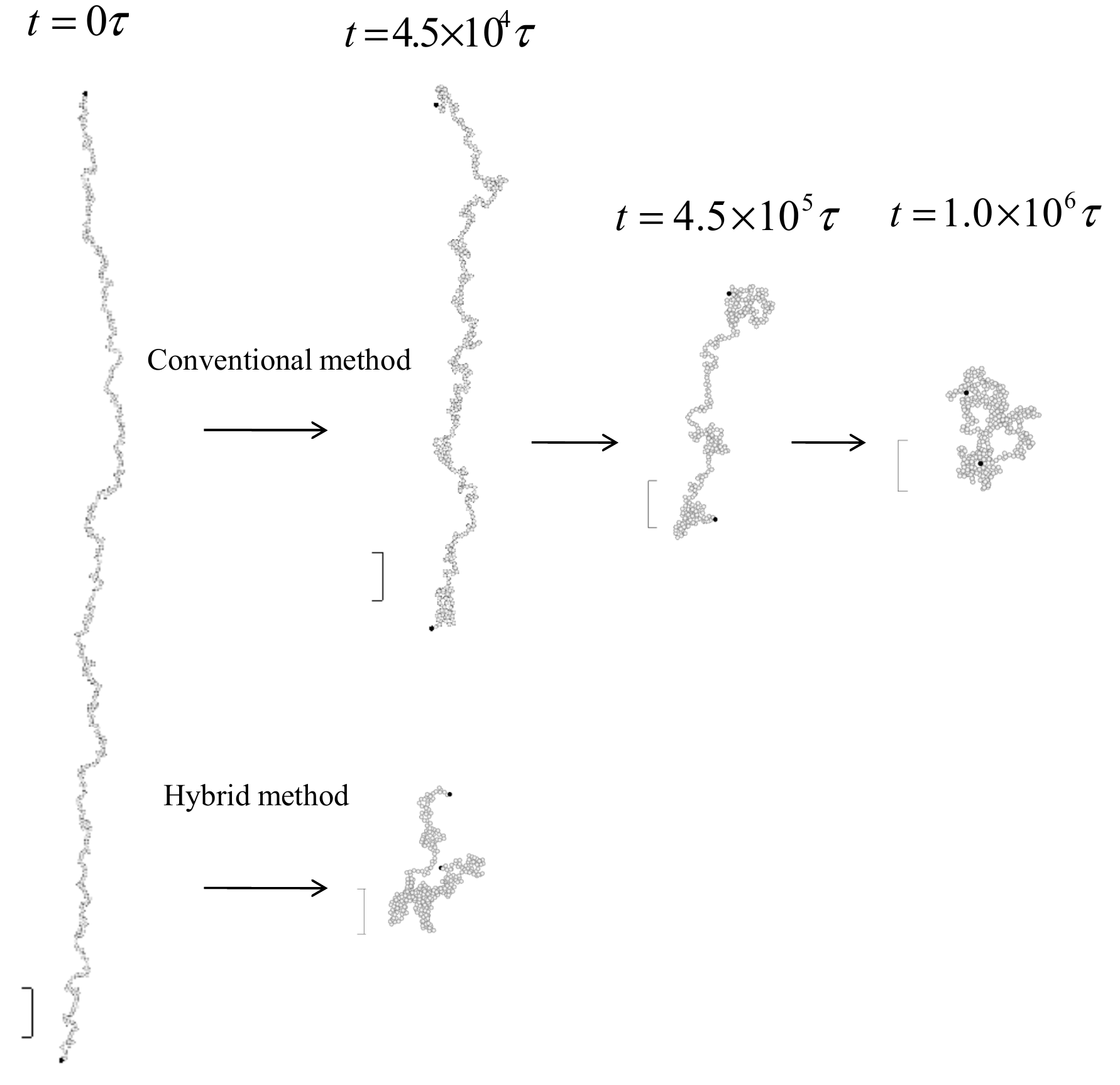}
\end{center}
\caption{ Time convolution of a single polymer configuration obtained
  by the conventional method, compared to the hybrid method.  This is one
  polymer chosen from ten polymers in the simulation system.  Guides
  in figure indicate the length of 10 $\sigma$.  }
\label{config}
\end{figure}

Snapshots of a single polymer obtained by the conventional method and
by the hybrid method are shown in Fig. \ref{config}.  We start the MD
calculations using the same initial configuration of a long stretched
chain as shown in Fig. \ref{config}($t=0$).  At $t=4.5\times 10^{4}
\tau $, the polymer configuration of the conventional method still has a
stretched chain structure, while that of the hybrid method has a well
equilibrated, entangled, structure.  In general a flexible polymer such as the
present polymer model has an entangled structure in the equilibrium
state.  A well-equilibrated state can be obtained at $t=4.5\times
10^{4} \tau $ by the hybrid method, while we manage to obtain
the equilibrium structure only at $t \sim 10^{6} \tau $ by the conventional
method.

\begin{figure}
\begin{center}
\includegraphics[width=0.6\columnwidth]{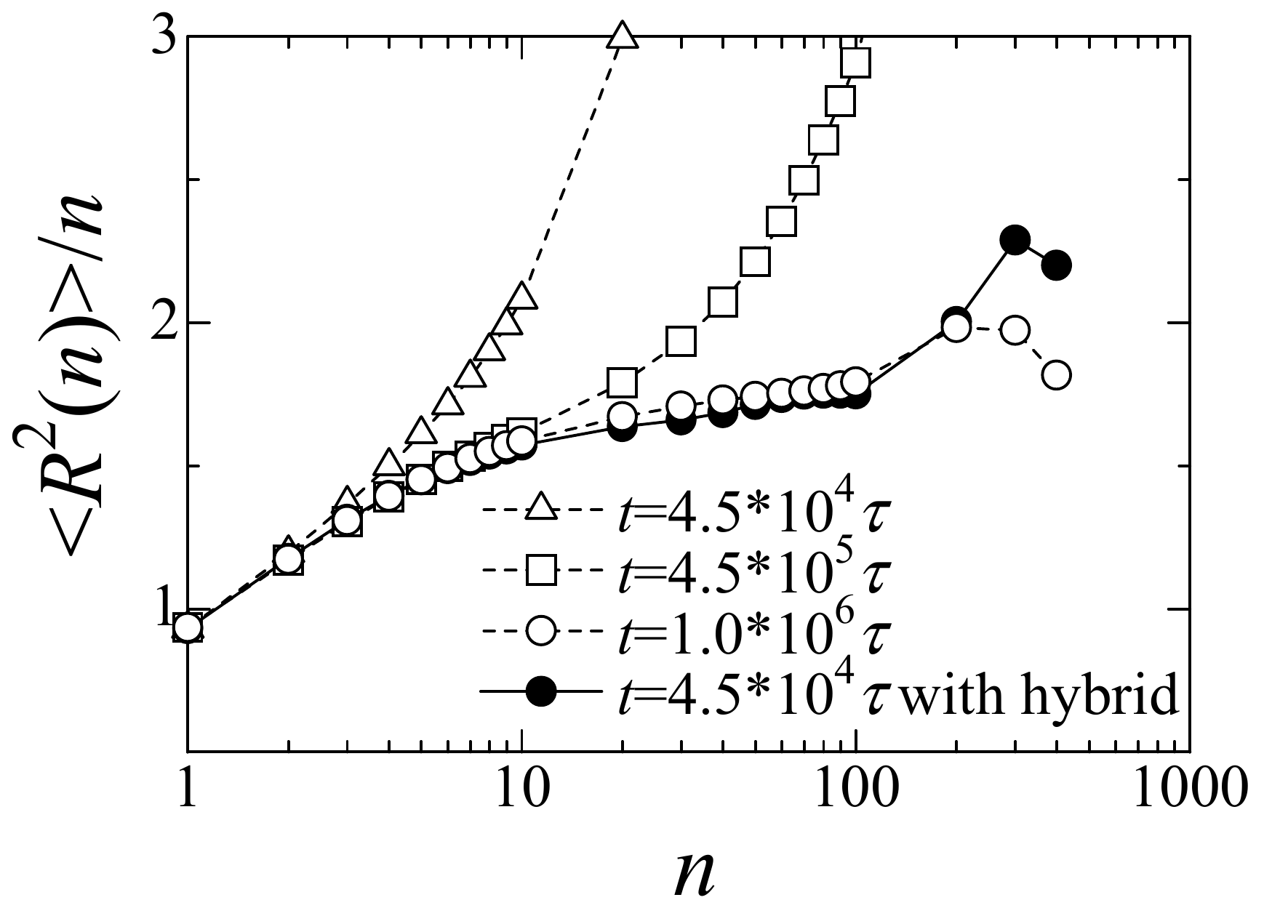}
\end{center}
\caption{ Mean square internal distances obtained by the conventional
  method (broken lines) and hybrid method (solid line).  For the
  conventional method, results at three different times ($t=4.5 \times
  10^{4} \tau, 4.5 \times 10^{5} \tau$ and $1.0 \times 10^{6} \tau $) are shown.
}
\label{re}
\end{figure}

In addition, we derive the mean square internal distance $<R(n)^{2}>$,
averaged over all internal distances $n=|i-j|$ along all the polymer
chains, where $i<j\in[1,N]$ are the monomer indices.  
It is also confirmed from the time convolution of $<R(n)^{2}>$ that 
the MD calculation is accelerated toward equilibrium state by hybrid method.  
It is shown in Fig.\ref{re} that the curve of $<R(n)^{2}>/n$ obtained 
by the hybrid method at $t=4.5 \times 10^{4} \tau $ is consistent with that of the
final equilibrium state.  The curve obtained by the conventional
method at $t=4.5 \times 10^{4} \tau $ is far from the equilibrium, and
even at a simulation time of $t=4.5 \times 10^{5} \tau $, it still
does not converge to the equilibrium.  
$<R(n)^{2}>$ for long distances $(n>200)$ have statistical errors due to the small number of polymers
(10 polymer chains in the present calculation).
As the $n$ reaches chain length $N$, there exist less and less pairs of monomers,
and larger statistical errors.  
These errors are enhanced by the small number of polymer chains 
and are thus influencing the result for large $n$.

\section{Discussion}
\begin{figure}
\begin{center}
\includegraphics[width=0.6\columnwidth]{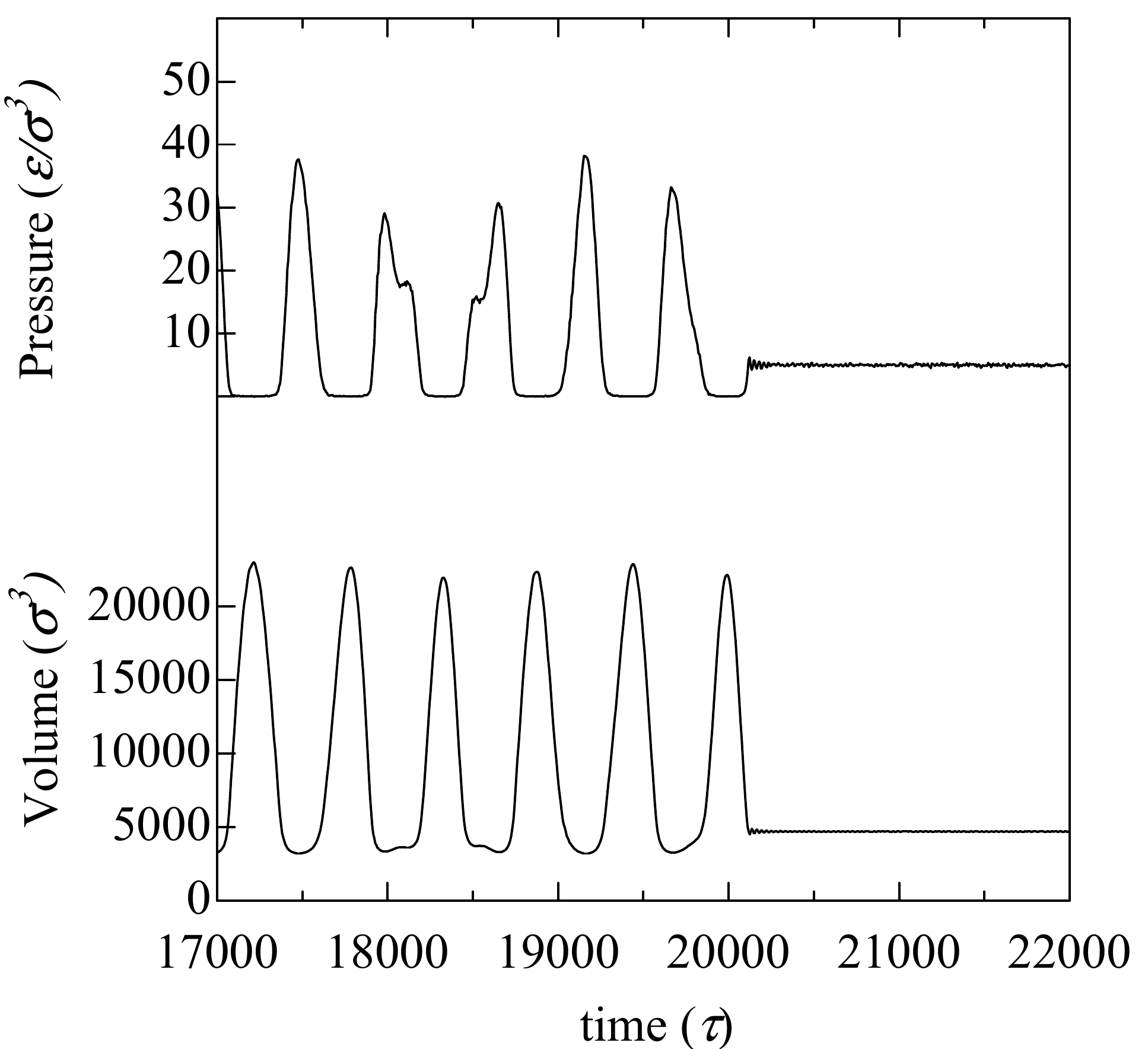}
\end{center}
\caption{ Time convolution of the internal pressure and the volume of
  the MD cell.  The calculation by the hybrid method stops at $t=20000
  \tau $, and the constant-pressure MD calculation is continued after
  that.  }
\label{p_v}
\end{figure}

The fluctuation of the end-to-end distance $R$ of the hybrid model means
that the chain structure of the polymer is fluctuating wildly during
the simulation.  The time convolution of the internal pressure of the CG
polymer system and the volume of the MD cell is shown in
Fig. \ref{p_v}.  It can be seen in Fig. \ref{p_v} that the internal
pressure and the system volume are widely fluctuating.  It can also
be seen that the time convolution of the volume is out of phase with
 the internal pressure.  These large-scale
fluctuations arise in the polymer system; low pressure induces a 
large expansion of the volume, and a large pressure compresses
the system.  Hence the number density of the system is also widely
fluctuating.

\begin{figure}
\begin{center}
\includegraphics[width=0.6\columnwidth]{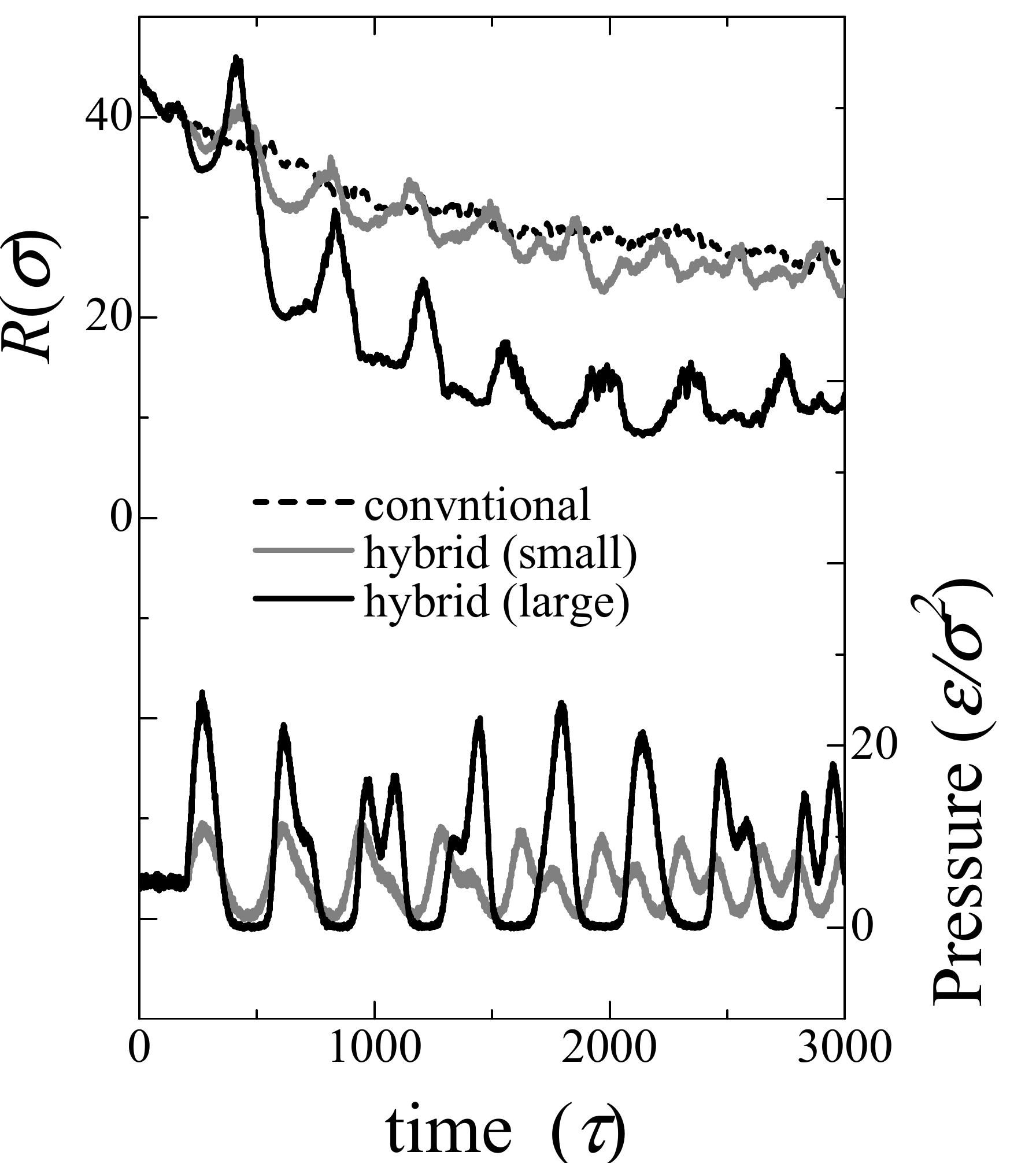}
\end{center}
\caption{ Time convolution of the end-to-end distance $R$ and the
  internal pressure.  The results are obtained with a smaller system
  than in the present calculations, where a single polymer consists of 100 coarse-grain monomers.  
  The gray line indicates the result obtained in the case of a small fluctuation by the hybrid method, 
  while the solid line indicates the result in the case of large-scale fluctuations.  }
\label{endtoend100}
\end{figure}

These large fluctuations in the polymer system are generated by the
vibrations of the springs.  The motion of the springs is connected
with that of the MD cell as is described by Equation \ref{eq8}.  The
large-scale dynamics of the elastic continuum leads to the fluctuation of
the CG polymer system and results in the fast convergence to the
equilibrium.  If only small stress of the elastic continuum acts on
the CG polymer system, the convergence is not improved.  It is
confirmed in Fig. \ref{endtoend100} that a small vibration of the springs
induces small fluctuations in the polymer system, and the convergence is not
improved compared to the conventional method in this case.

\begin{figure}
\begin{center}
\includegraphics[width=0.6\columnwidth]{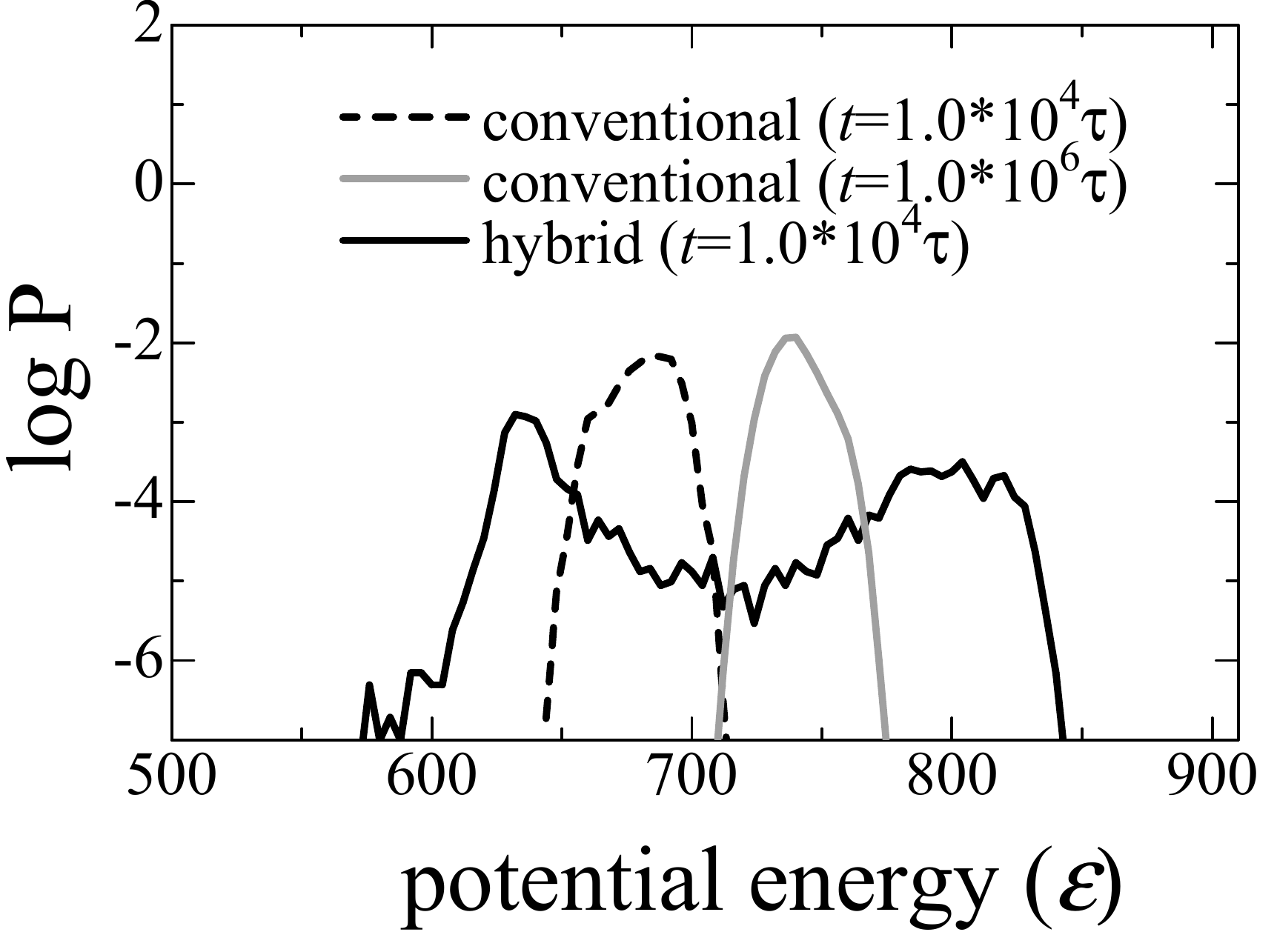}
\end{center}
\caption{ The probability distribution functions of the bending
  potential for the hybrid method together with the conventional
  method.  }
\label{pdf}
\end{figure}

Due to the large-scale fluctuations, the trajectory of the polymer spreads
over a wide phase space and a wide range of energy.  The present CG
polymer model has a bending potential, which is closely associated
with the chain structure and its flexibility.  The
probability distribution function of the bending potential is shown in
Fig. \ref{pdf}.  
In the convensional method, an early state ($t=1.0 \times 10^{4} \tau$) 
is trapped in a limited local energy range around $690\epsilon$, 
which is separated from final equilibrium states of the energy around $740\epsilon$.
In contrast, a much wider energy range is coverd by the hybrid method. 
This suggests that the hybrid method more easily allows us to overcome energetic
barriers separating the initial state from the final equilibrium state.
A large variety of chain structures is generated and its trajectory is
over a much wider phase space than that of the conventional method.

In our previous studies\cite{kim2007,senda2009}, a one-dimensional model
was calculated by the hybrid method and the large-scale fluctuation
causes the generation of a variety of phonons in the particle system.
The phonons obtained by the hybrid model reproduced those by large-scale
all-atom calculations.  It was shown that the hybrid model enables us
to extend the {\it spatial scale} to much larger values.  In the present study,
the hybrid method is applied to the long-chain polymer that has very slow diffusion dynamics.  The required simulation time to reach
the equilibrium state of the polymer melt is reduced drastically.  The
present hybrid model thus enables us to reach a much wider {\it time scale}
than that by the conventional MD method.

Although the present model is simple, the hybrid method can be applied
to more realistic and complex systems such as polycarbonate
\cite{karen2007} and protein molecules.  The equilibrium structure and
various properties at finite temperatures can be obtained with
reasonable computational cost.

\section{Conclusion}
We couple a CG polymer model and an elastic continuum using the hybrid
method.  The polymer melt consisting of the CG polymers is simulated
and it is shown that fast convergence towards the equilibrium state of
the polymer melt is achieved.  The elastic continuum of the hybrid
model acts on the polymer system and produces large scale
fluctuations. The fluctuations allow the polymer system to sample over
a much wider phase space than the conventional method, inducing a variety
of polymer states, and leads to fast convergence toward the
equilibrium.  
   
\section{Acknowledgements}
We thank Dr Karen Johnston for helpful discussion and fruitful collaboration. 
This work was supported by Functional Materials Programme of Tekes, Finland, 
and by a Grant-in-Aid for Scientific Research in Priority Areas (no.17064012) 
from the Ministry of Education, Culture, Sports, Science and Technology, Japan.
We also thank Japan Aerospace Exploration Agency (JAXA) 
for allowing us to use JAXA supercomputer system.

\end{document}